\documentclass{IEEEtaes}

\usepackage{color,array,amsthm}
\usepackage{graphicx}
\usepackage{cite}
\usepackage{amsmath,amssymb,amsfonts}
\usepackage{graphicx}
\usepackage{textcomp}
\usepackage{xcolor}
\usepackage{booktabs}
\usepackage{multirow}
\usepackage{lipsum}
\usepackage{hyperref}

\usepackage[capitalise,noabbrev,nameinlink]{cleveref}
\Crefname{subsection}{Subsection}{Subsections}
\crefname{subsection}{Subsection}{Subsections}

\usepackage[makeroom]{cancel}

\usepackage[acronyms,nonumberlist,nopostdot,nomain,nogroupskip,acronymlists={hidden}]{glossaries} 
\newglossary[algh]{hidden}{acrh}{acnh}{Hidden Acronyms}

\usepackage{algorithm}
\usepackage[noend]{algpseudocode}
\algrenewcommand\algorithmicindent{1em} 

\jvol{XX}
\jnum{XX}
\jmonth{XXXXX}
\paper{1234567}
\pubyear{2024}
\doiinfo{TAES.2024.Doi Number}

\usepackage{caption}

\newacronym{mmwave}{mmWave}{millimeter-wave}
\newacronym{rcs}{RCS}{Radar Cross Section}
\newacronym{ratr}{RATR}{Radar Automated Target Recognition}
\newacronym{atr}{ATR}{Automated Target Recognition}
\newacronym{sar}{SAR}{Synthetic Aperture Radar}
\newacronym{sbr}{SBR}{Shooting and Bouncing Ray}
\newacronym{rnn}{RNN}{Recurrent Neural Network}
\newacronym{cnn}{CNN}{Convolutional Neural Network}
\newacronym{lstm}{LSTM}{Long Short Term Memory}
\newacronym{mds}{MDS}{Micro-Doppler Signature}
\newacronym{mdf}{MDF}{Micro-Doppler Frequency}
\newacronym{uav}{UAV}{Unmanned Aerial Vehicle}
\newacronym{rlos}{RLOS}{Radar Line Of Sight}
\newacronym{isar}{ISAR}{Inverse Synthetic Aperature Radar}
\newacronym{gru}{GRU}{Gated Recurrent Unit}
\newacronym{ffnn}{FFNN}{Feed Forward Neural Network}
\newacronym{mlp}{MLP}{Multi Layer Perceptron}
\newacronym{snr}{SNR}{Signal-to-Noise-Ratio}
\newacronym{acgn}{ACGN}{Additive Colored Gaussian Noise}
\newacronym{slurm}{SLURM}{Simple Linux Utility for Resource Management}
\newacronym{ml}{ML}{Machine Learning}
\newacronym{hrrp}{HRRP}{High Range Resolution Profile}
\newacronym{rbe}{RBE}{Recursive Bayesian Estimation}
\newacronym{hmm}{HMM}{Hidden Markov Model}
\newacronym{svm}{SVM}{Support Vector Machine}
\newacronym{nctr}{NCTR}{Non Cooperative Target Recognition}
\newacronym{gmm}{GMM}{Gaussian Mixture Model}
\newacronym{gnn}{GNN}{Graph Neural Network}
\newacronym{rbc}{RBC}{Recursive Bayesian Classification}
\newacronym{obf}{OBF}{Optimal Bayesian Fusion}
\newacronym{nn}{NN}{Neural Network}
\newacronym{tx}{TX}{Transmitter}
\newacronym{rx}{RX}{Receiver}
\newacronym{selx}{SELX}{SELection of tX and rX}
\newacronym{ew}{EMW}{Electromagnetic Wave}
\newacronym{xgboost}{XGBoost}{Extreme Gradient Boosting}
\newacronym{dl}{DL}{Deep Learning}
\newacronym{knn}{KNN}{K - Nearest Neighbors}
\newacronym{mle}{MLE}{Maximum Likelihood Estimation}

\begin{document}

\title{Multistatic-Radar RCS-Signature Recognition of Aerial Vehicles: A Bayesian Fusion Approach}
% \title{Optimal Bayesian Target Recognition for Unmanned Aerial Vehicles Based on Radar Cross Section} 

\author{Michael Potter}
\member{Student Member, IEEE}
\affil{Northeastern University, Boston, MA 02115, USA} 

\author{Murat Akcakaya}
\member{Senior Member, IEEE}
\affil{University of Pittsburgh, Pittsburgh, PA 15260, USA} 

\author{Marius Necsoiu}
\member{Member, IEEE}
\affil{DEVCOM ARL, San Antonio, TX 78204, USA} 

\author{Gunar Schirner}
\member{Member, IEEE}
\affil{Northeastern University, Boston, MA 02115, USA}

\author{Deniz Erdo\u{g}mu\c{s}}
\member{Senior Member, IEEE}
\affil{Northeastern University, Boston, MA 02115, USA} 

\author{Tales Imbiriba}
\member{Member, IEEE}
\affil{Northeastern University, Boston, MA 02115, USA} 
%% \author{FOURTH D. AUTHOR}
%% \affil{University of Colorado, Colorado, USA}

\receiveddate{{\tiny Manuscript received February 29, 2024; revised May 31, 2024; accepted August 06, 2024.} \\ \\ {\tiny
Research was sponsored by the Army Research Laboratory and was accomplished under Cooperative Agreement Number W911NF-23-2-0014. The views and conclusions contained in this document are those of the authors and should not be interpreted as representing the official policies, either expressed or implied, of the Army Research Laboratory or the U.S. Government. The U.S. Government is authorized to reproduce and distribute reprints for Government purposes notwithstanding any copyright notation herein.} \\ \\
{\tiny © 2024 IEEE: Personal use of this material is permitted. Permission from IEEE must be obtained for all other uses, in
any current or future media, including reprinting/republishing this material for advertising or promotional purposes, creating
new collective works, for resale or redistribution to servers or lists, or reuse of any copyrighted component of this work in
other works. Submitted for review to: IEEE Journal - Transactions on Aerospace and Electronics - 2024}
}

\markboth{POTTER ET AL.}{Multistatic-Radar RCS-Signature Recognition of Aerial Vehicles: A Bayesian Fusion Approach}
\maketitle

\begin{abstract} \gls{ratr} for \glspl{uav} involves transmitting \glspl{ew} and performing target type recognition on the received radar echo, which has important applications in defense and aerospace. Previous work has demonstrated the benefits of employing multistatic radar configurations in \gls{ratr} compared to monostatic radar configurations. However, multistatic radar configurations commonly use fusion methods which combine the classification vectors of multiple individual radars suboptimally from a probabilistic perspective. 

To address this issue, this work leverages Bayesian analysis to provide a fully Bayesian \gls{ratr} framework for \gls{uav} type classification. Specifically, we employ an \gls{obf} method, from the Bayesian perspective of expected 0-1 loss, to formulate a posterior  distribution that aggregates the classification probability vectors from multiple individual radar observations at a given time step. This \gls{obf} method is used to update a separate \gls{rbc} posterior distribution on the target \gls{uav} type. The \gls{rbc} posterior is conditioned on all historical observations made from multiple radars across multiple time steps. 

To evaluate the proposed approach, we simulate random walk trajectories for seven drones and correspond the target's aspect angles to \gls{rcs} measurements acquired in an anechoic chamber. We then compare the performance of single radar \gls{atr} system and suboptimal fusion methods against the \gls{obf} method. We empirically show that the \gls{obf} method, integrated with \gls{rbc}, significantly outperforms other fusion methods and single radar configuration in terms of classification accuracy.
\\ \\
Code: \href{https://github.com/mlpotter/RCS_ATR}{https://github.com/mlpotter/RCS\_ATR}
\end{abstract}

\vspace{-1em}

\glsresetall

\section{INTRODUCTION}
\label{sec:intro}

\gls{ratr} technology has revolutionized the domain of target recognition across space, ground, air, and sea-surface targets \cite{jiang2023radar}. Radar advancements have facilitated the extraction of detailed target feature information, including \gls{hrrp}, \gls{sar}, \gls{isar}, \gls{rcs} and Micro-Doppler frequency \cite{eaves2012principles}. These features have enabled \gls{ml} and \gls{dl} models to outperform traditional hand-crafted target recognition frameworks \cite{jiang2023radar,cai2021machine}. Furthermore, \gls{ratr} seamlessly integrates with various downstream defense applications such as weapon localization, ballistic missile defense, air surveillance, ground and area surveillance among others \cite{8341898}. While \gls{ratr} is applied to various target types, our focus will center on the recognition of \glspl{uav}. Henceforth, the terms \glspl{uav} and drones will be used interchangeably.

\glspl{uav} are a class of aircraft that do not carry a human operator and fly autonomously or are piloted remotely \cite{knuthwebsite}. Historically, \glspl{uav} were designed solely for military applications such as surveillance and reconnaissance, target acquisition, search and rescue, and force protection \cite{lewis2011drones,mahadevan2010military}. Recently we have entered the ``\textit{Drone Age}''
\cite{beesley2023head}, aka personal \textcolor{black}{\gls{uav}} area; where there has been a rapid increase in civilian \gls{uav} use for cinematography, tourism, commercial ads, real estate surveying and hot-spot / communications \cite{civiliandronesurvey,yaacoub2020security}.

However, the potential misuse of \glspl{uav} poses significant security and safety threats, which is prompting governmental and law enforcement agencies to implement regulations and countermeasures \cite{bassi2020here,madiega2021artificial}. Adversaries may transport communication jammers with \glspl{uav} \cite{securityintelligence}, perform cyber-physical attacks on \glspl{uav} (possibly with kill-switches)\cite{cyberphysical}, or conduct espionage via video streaming from \glspl{uav} \cite{mekdad2023survey}. Criminals may perform drug smuggling, extortion(with captured footage), and cyber attacks (on short-range Wi-Fi, Bluetooth, and other wireless devices) \cite{yaacoub2020security}. These  security and safety issues are exacerbated by the increased presence of unauthorized and unregistered \glspl{uav} \cite{unregistereduav}. The growing concerns of the new Drone Age, coupled with the increasing nefarious use of \glspl{uav} \cite{markarian2020countermeasures}, necessitates robust, accurate, and fast \gls{ratr} frameworks.

Radar systems transmit \glspl{ew} directed at a target, and receiving the reflected \glspl{ew} (aka radar echoes) \cite{richards2010principles}. Radar systems are ubiquitous in \gls{atr} because of the long-range detection capability, ability to penetrate obstacles such as atmospheric conditions, and versatility in capturing detailed target features \cite{richards2010principles}. Popular radar-based methods of generating target features are \gls{isar}, Micro-Doppler signature, and \gls{rcs} signatures. \gls{isar} generates high resolution radar images by using the Fourier transform and explotation of the relative motion of the target \gls{uav} to create a larger ``synthetic'' aperture \cite{chen2014inverse}. However, \gls{isar} images degrade significantly when the target \gls{uav} has complex motion, such as non-uniform pitching, rolling, and yawing \cite{Ezuma2}.  Micro-Doppler signatures are the frequency modulations around the main Doppler shift due to the \gls{uav} containing small parts with additional micro-motions, such as the blade propellers of a drone \cite{clemente2013developments}. However, Micro-Doppler signatures may vanish due to the relative orientation of the drone. Both \gls{isar} and Micro-Doppler require high radar bandwidth for processing and generating target profiles, which is computationally expensive \cite{klaer2020investigation,chen2019micro}. The \gls{rcs} of a \gls{uav}, measured in dB m\textsuperscript{2} and proportionate to the received power at the radar, signifies the theoretical area that intercepts incident power; if this incident power were scattered isotropically, then it would produce an echo power at the radar equivalent to that of the actual \gls{uav} \cite{rcsdefinition}. \gls{rcs} signatures have low transmit power requirements, low bandwidth requirements, low memory footprint, and low computational complexity \cite{ehrman2010using,Ezuma2}. For these reasons, much of the literature focuses on \gls{rcs} data collected from transmitted signals with varying carrier frequencies. %We use the preprocessed \gls{rcs} signatures as input features to discriminate \gls{ml} classification models. 

Most of the \gls{uav} \gls{atr} literature using \gls{rcs} signatures focuses on monostatic radar configurations and does not utilize multiple individual monostatic radar observations at a single time step. However, there are common multistatic radar configuration fusion rules which are based on domain-expert knowledge.  The \gls{selx} fusion rule \cite{6136010,6212164} combines the classification probability vectors from each radar channel via a weighted linear combination, where  the weights are higher for the channels with the best range resolution and high \gls{snr}. The common \gls{snr} fusion rule \cite{derham2010ambiguity} also combines the classification probability vectors for each radar channel via a weighted linear combination, but the weights proportional to the \gls{snr} (higher \gls{snr} leads to larger weight). Other works employ heuristic soft-voting, which averages the probability vectors from each radar \cite{meng2022spatio}. However, domain expert's heuristics may be influenced by cognitive bias or limited by subjective interpretation of experts \cite{koehler2002calibration,yu2014decision}, leading to incorrect conclusions. Furthermore, Bayesian decision rules are known to be optimal decision-making strategies under uncertainty, compared to domain-expert's heuristic-based decisions \cite{smith2010bayesian,baron2014heuristics}. 

\textbf{Our contributions of this work:} To the best of our knowledge, this work is the first to provide a fully Bayesian \gls{ratr} framework for \gls{uav} type classification from \gls{rcs} time series data. We employ an \gls{obf} method, which is the Bayesian fusion method for optimal decisions with respect to 0-1 loss, to formulate a posterior  distribution from multiple individual radar observations at a given time step. This optimal method is then used to update a separate \gls{rbc} posterior distribution on the target \gls{uav} type, conditioned on all historical observations  from multiple radars over time. 

Thus, our approach facilitates efficient use of training data, informed classification decisions based on Bayesian principles, and enhanced robustness against increased uncertainty. Our method demonstrates a notable improvement in classification accuracy, with relative percentage increases of 35.71\% and 14.14\% at an \gls{snr} of 0 dB compared to single radar and \gls{rbc} with soft-voting methods, respectively. Moreover, our method achieves a desired classification accuracy (or correct individual prediction) in a shorter dwell time compared to single radar and \gls{rbc} with soft-voting methods.

The paper is structured as follows: Section II describes the related work for \gls{ratr} on \gls{rcs} data; Section III details the data, data simulation and dataset structure; Section IV explains the \gls{obf} method and \gls{rbc} for our \gls{ratr} framework; Section V discusses the experiment configuration and results; Section VI derives conclusions on our findings.
\vspace{-1em}

\section{RELATED WORK}
\label{sec:related_works}
We sort the literature on \gls{ratr} target recognition models into two types: statistical \gls{ml} and \gls{dl}. 

\textbf{Statistical \gls{ml}}: 
Many methods leverage generative probabilistic models using Bayes Theorem to classify a target \gls{uav} based on the highest class conditional likelihood  \cite{ezuma2021comparative,Ezuma2,4526420}. The method in \cite{ezuma2021comparative,Ezuma2} classified commercial drones with generative models such as Swerling (1-4), Gamma, \gls{gmm} and Naive Bayes on multiple independent \gls{rcs} observations simulated from an monostatic radar in an anechoic chamber. However, there could potentially exist a hidden temporal relationship among the \gls{rcs} observations. In the study of \cite{7511858}, unspecified aircrafts were classified based on \gls{rcs} time series data using an \gls{hmm} fitted with the Viterbi algorithm. In this model, the hidden state represented the contiguous angular sector of the \gls{uav}-radar orientation, with the observations comprising \gls{rcs} measurements. While generative models are designed to capture the joint distribution of both data and class labels, this approach proves computationally inefficient and overly demanding on data when our primary goal is solely to make classification decisions \cite{bishop2006pattern}.

\begin{figure*}[ht!]
    \centering
    \includegraphics[width=14.8cm]{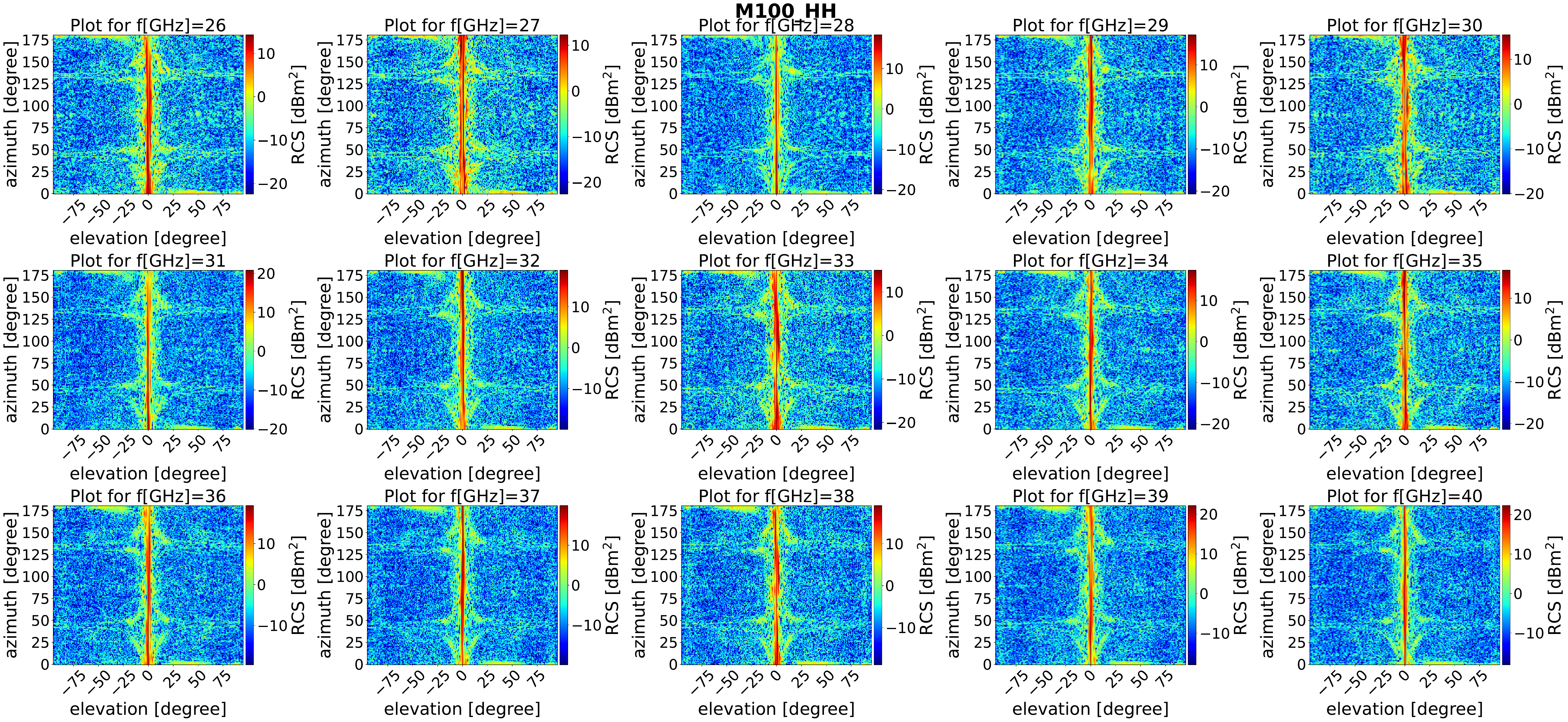}
    \caption{\small M100 HH polarization \gls{rcs} signature images from the Indoor Near Field experiments from \cite{rcs_data}}
    \label{fig:M100_label}
    \vspace{-2em}
\end{figure*}

Discriminative \gls{ml} models directly find the posterior probability on the \gls{uav} type given \gls{rcs} data, and have been used in \gls{ratr} for \gls{uav} type classification \cite{ezuma2021comparative,10242797,9391260,6875676}. The study described in \cite{6875676} employed summary statistics of \gls{rcs} time series data to classify between UAV and non-UAV tracks, utilizing both \gls{mlp} and \gls{svm} models. However, Ezuma et al showed that Tree-based classifiers outperformed generative, other discriminative \gls{ml} models (such as \gls{svm}, \gls{knn}, and ensembles), and deep learning models  \cite{ezuma2021comparative}. Furthermore, \cite{9266405} showed Random Forests significantly outperform \gls{svm} when classifying commercial drones using micro-Doppler signatures derived from  simulated \gls{rcs} time series data \cite{9391260}. While these methods demonstrate high classification accuracy in monostatic radar experiments, they do not take advantage of or account for multistatic radar configurations. Instead of consolidating all the features from multiple RCS observations into a large input space, it is more effective to construct individual models for each time slice or radar viewpoint \cite{bishop2006pattern}. We now shift our focus to \gls{dl} methods, acknowledging the undeniable ascent of \gls{dl} \cite{Goodfellow-et-al-2016,deng2018artificial}.

\textbf{\gls{dl}}: The enhanced capability of radar to extract complex target features has sparked an increasing demand for models capable of using complex data, exemplified by the rising use of \gls{dl} model \cite{derham2010ambiguity,jiang2023radar}. The work in \cite{wengrowski2019deep} was the first application of \gls{cnn} for monostatic radar classification based on simulated \gls{rcs} time series data of geometric shapes. However, \glspl{cnn} do not incorporate long-term temporal dynamics into their modeling approach. To address this limitation and capture long-term and short term temporal dynamics, many studies have used \glspl{rnn} and \glspl{lstm} networks \cite{bishop2023deep}. These works classify a sliding time window or the entire trajectory of \gls{rcs} time series data corresponding to monostatic radars tracking an \gls{uav} \cite{mansukhani2021rcs,sehgal2019automatic,fu2021deep}. \cite{zhu2020rcs} combined a bidirectional \gls{gru} \gls{rnn} and a \gls{cnn} to extract features from \gls{rcs} time series data of geometric shapes and small-sized planes, subsequently using an \gls{ffnn} for classification. Lastly, \cite{meng2022spatio} employed a spatio-temporal-frequency \gls{gnn} with \gls{rnn} aggregation to classify two simulated aircraft with different micromotions tracked by a synthetic heterogeneous distributed radar system. While numerous \gls{dl} approaches have been utilized in \gls{rcs} based \gls{ratr} classification frameworks, challenges still persist, including the necessity for substantial data to generalize to unseen data \cite{bansal2022systematic}, overconfident predictions \cite{immer2021improving}, and a lack of interpretability \cite{castelvecchi2016can}. Furthermore, Ezuma et al showed that Tree-based \gls{ml} models outperformed traditional deep complex \gls{nn} architectures for commercial drone classification. Furthermore \cite{shwartz2022tabular} showed that for tabular datasets that \gls{dl} does not perform as well as \gls{ml}. Thus this paper focuses on the following \gls{ml} models: \gls{xgboost}, \gls{mlp} with three hidden layers, and logistic regression. 

\vspace{-1em}
\section{Multi-Static RCS Dataset Generation}
\label{sec:data_gen}
This section outlines the process of mapping simulated azimuth and elevation angles to \gls{rcs} data measured in an anechoic chamber. This involves calculating the target \gls{uav} aspect angles (azimuth and elevation) relative to the \gls{rlos} while considering the target \gls{uav} pose (yaw $\gamma$, pitch $\alpha$, roll $\beta$ and translation $T$). The pose of the target \gls{uav} will change over time as a function of a random walk kinematic model.

\begin{figure}[h!]
    \centering
    \includegraphics[width=6.0cm]{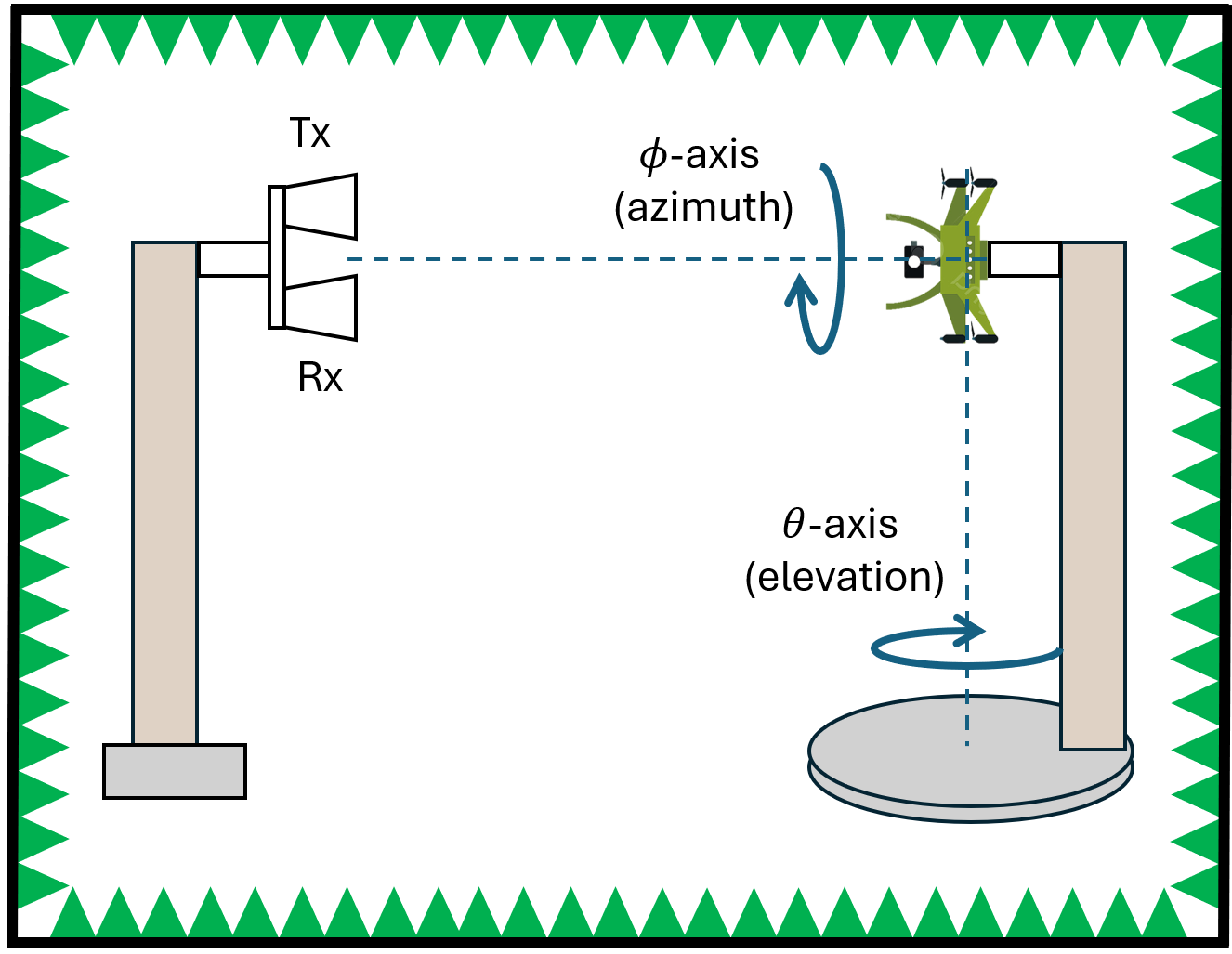}
    \caption{\small Indoor Near Field \gls{rcs} signature collection utilizing different polarization combinations of \gls{ew} signals transmitted and received from from quasi-monstatistc radar.}
    \label{fig:anechoic_chamber}\
    \hspace{-1em}
\end{figure}

\subsection{Existing Data}
\label{sec:existdata}
We analyze the approximate \gls{rcs} signatures of 7 different \glspl{uav} collected in \cite{rcs_data}: the F450, Heli, Hexa, M100, P4P, Walkera, and Y600 .  The data was collected in an anechoic chamber, where a quasi-monostatic radar transmitted \glspl{ew} at $F=15$ frequencies (26-40 GHz in 1 GHz steps) while stepped motors rotated the \gls{uav} around the azimuth axis ($\phi$ in \cref{fig:anechoic_chamber}) and the elevation axis ($\theta$ in \cref{fig:anechoic_chamber}) \cite{rcs_data}. \textcolor{black}{We note that \glspl{uav} were non-operational during the data collection process, and therefore the measured \gls{rcs} signature of the drones in flight may slightly change due to the rotating propellers.} The azimuth spans from 0 to 180 degrees, while the elevation extends from -95 to 95 degrees, both in 1-degree increments. An example of the \gls{rcs} image of a \gls{uav} from \cite{rcs_data} is shown in \cref{fig:M100_label}. We leverage the real \gls{rcs} signatures to create training and testing datasets corresponding to simulated/synthetic radar locations and drone trajectories.

\vspace{-1em}
\subsection{Training Data Generation}
\label{sec:traindata}
We sample azimuth and elevation uniformly at random, where the spread is bounded by each drone's minimum and maximum aspect angles from the experimental setup in \cite{rcs_data}. The azimuth and elevation samples, $\phi$ and $\theta$ respectively, are mapped to a measured \gls{rcs} signature $\sigma \in \mathcal{R}^{F}$ from \cite{rcs_data}, where we \textcolor{black}{bilinearly} interpolate the mapped \gls{rcs} measurements for continuous azimuth and elevations not collected in \cite{rcs_data}. For each \gls{uav} type $c \in \mathcal{C}_{train}=[1, \ldots, 7]$, we generate $N_{train}$ samples such that $|\mathcal{D}_{train}| = |\mathcal{C}_{train}|N_{train}$. Training a discriminative \gls{ml} classifier under the perspective of a single radar at a single time point enables efficient sampling and training. Generating training data for multistatic radar configurations would deal with the curse of dimensionality, where trajectories (a sample trajectory being multiple time points concatenated) compounded with multiple radars would exponentially increase the required number of samples to adequately cover the input space. In summary, the training dataset $\mathcal{D}_{train}$ has the following structure:

\vspace{-\baselineskip} 
\begin{align}
    \mathcal{D}_{train} = \{(x_i,c_i)\}_{i=1}^{N_{train}}
\end{align}
where $x_i^{(t)}=[\sigma_i^{(t)},\phi_i^{(t)},\theta_i^{(t)}]$ contains the noisy \gls{rcs} signature, the noisy azimuth and the noisy elevation. The noisy \gls{rcs} signature, azimuth, and elevation will be discussed in \cref{sec:data_gen}. \textcolor{black}{Note that the sampling approach of bilinear interpolation and then addition of independent Gaussian noise ensures unique \gls{rcs} measurements.}

\begin{figure}[h!]
\hspace*{0cm}                              
    \centering
    \includegraphics[width=9cm]{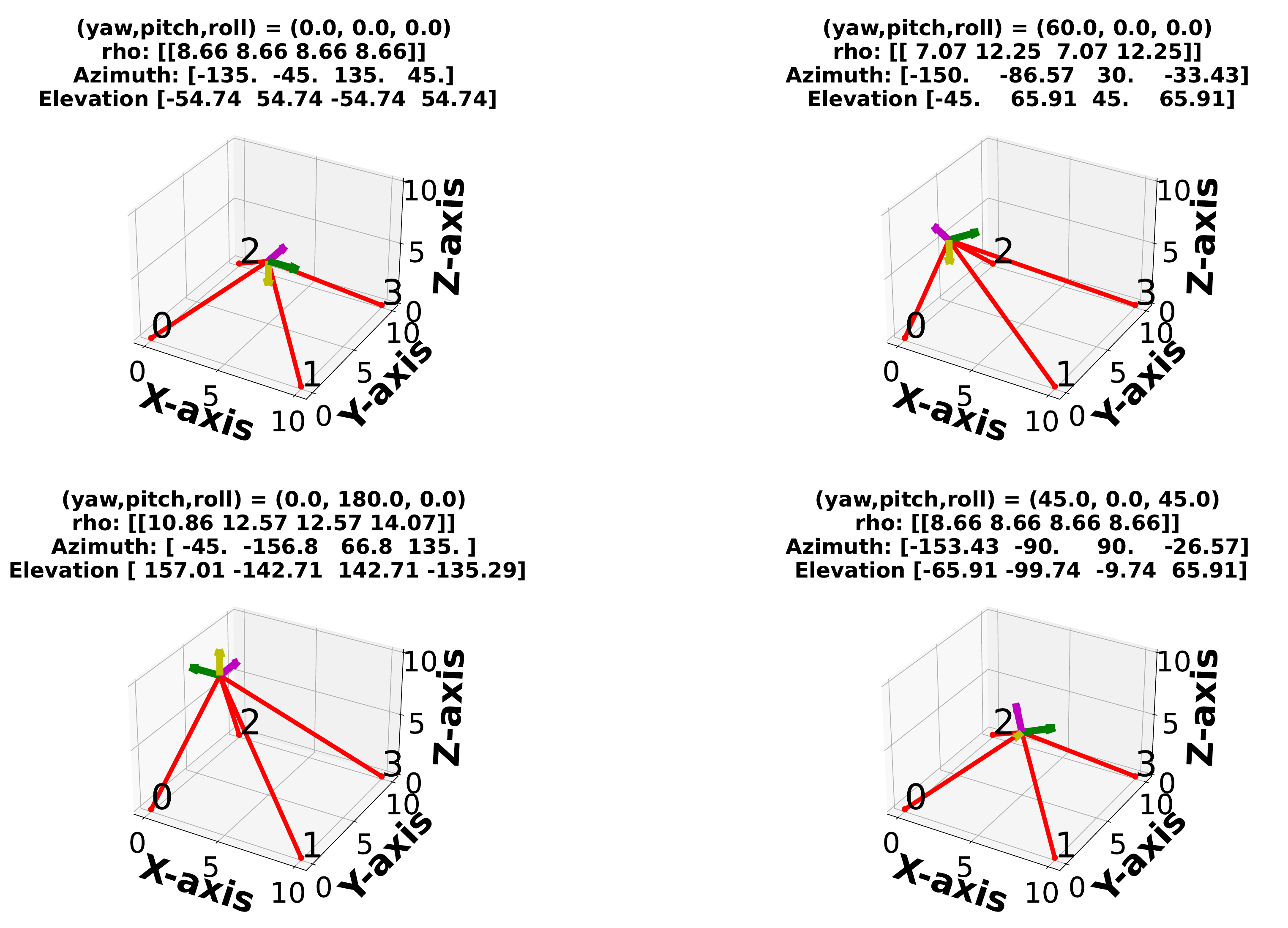}
    \caption{\small Example which illustrates the \gls{rlos} between multiple radars (4) and a single target. The radar positions are the red markers numbered (0-3). The x-axis, y-axis, and z-axis of the target \gls{uav} coordinate frame are the green, magenta, and yellow arrows respectively the target \gls{uav} position and coordinate frame is randomly simulated using translation, yaw, pitch, and roll homogeneous matrices.}
    \label{fig:simulated_rcs}
    \vspace{-1em}
\end{figure}

\begin{figure}[h!]
\hspace*{0cm}                              
    \centering
    \includegraphics[width=0.6\linewidth]{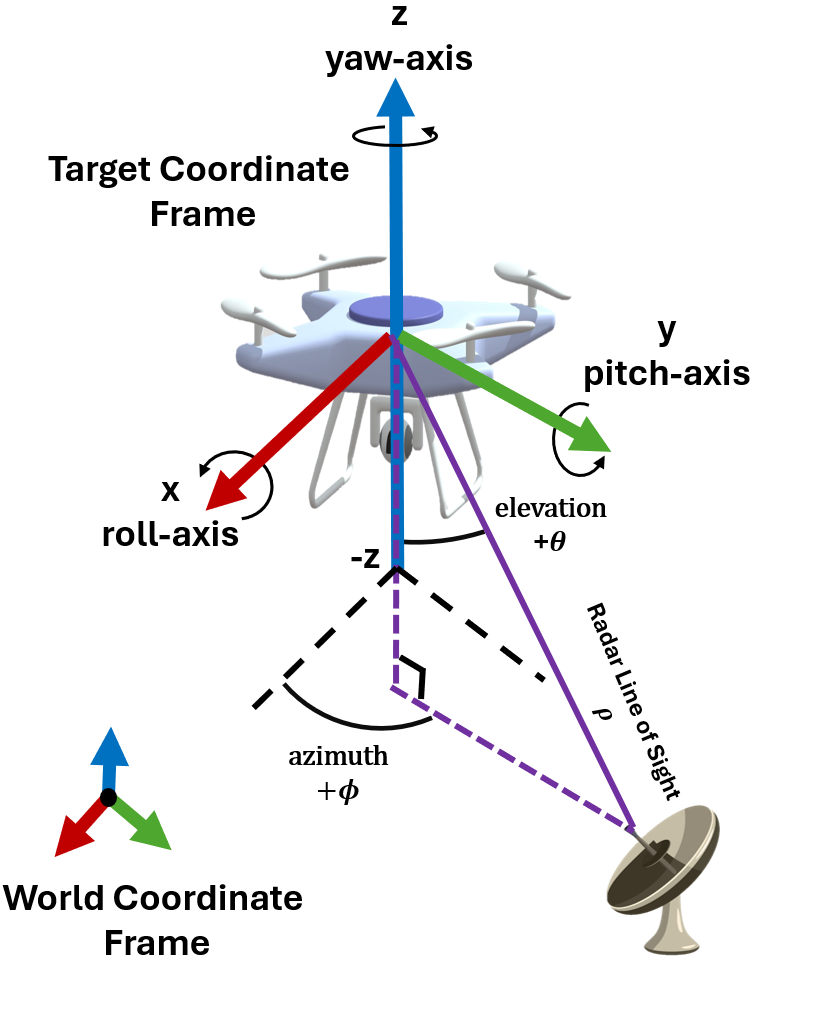}
    \caption{\small \textcolor{black}{The  relationship between the target coordinate frame, the radar and \gls{rlos}, and the world coordinate frame. The x-axis, y-axis, and z-axis of the target \gls{uav} coordinate frame are the red, green, and blue arrows respectively with  the x-axis as the forward direction. The position of the target is the translation with respect to the world coordinate frame.}}
    \label{fig:yawpitchrolldiagram}
    \vspace{-1em}
\end{figure}

\vspace{-1em}
\subsection{Testing Data Generation}
\label{sec:testdata}
We generate time series data of azimuth, elevation, and the corresponding \gls{rcs} signature by simulating multiple \gls{rlos} to a target \gls{uav} being tracked for $L=100$ time steps along random walk trajectories. For each \gls{uav} type $c \in \mathcal{C}_{test}$, we generate $N_{test}$ trajectories such that $|\mathcal{D}_{test}| = |\mathcal{C}_{test}|N_{test}$. The target \gls{uav} trajectory follows a kinematic model of a drone moving at constant velocity $v_x=50 [m/s]$ with random yaw and roll rotation jitters at a time resolution of $\Delta t = 0.1$ seconds: 
\begin{align}
    d^{(t)} &= R_{\gamma^{(t-1)}} R_{\alpha^{(t-1)}} R_{\beta^{(t-1)}}  e_x \\
    \mathcal{U}^{(t)} &= \mathcal{U}^{(t-1)}_d + v_x \Delta t d^{(t)} \\
    \gamma^{(t)} &= \gamma^{(t-1)} + \epsilon_{\gamma}; &&
    \epsilon_{\gamma} \sim \mathcal{N}(0^{\circ},144^{\circ}) \\
    \alpha^{(t)} &= 0  \\
    \beta^{(t)} &= \beta^{(t-1)} + \epsilon_{\beta}; &&
    \epsilon_{\beta} \sim \mathcal{N}(0^{\circ},81^{\circ}) 
\end{align}
where the positive $x$-axis of the target \gls{uav} coordinate frame is the \gls{uav} ``forward facing'' direction, $e_x$ is the x-axis unit vector, and $\mathcal{U}$ is the \gls{uav} position. \textcolor{black}{Furthermore, $R_{\gamma}, R_{\alpha}, R_{\beta}$ are the  Euler angle rotations for yaw, pitch, roll respectively of the target coordinate frame , and $d$ is the x-axis direction of the rotated target coordinate frame.} The  initial position and yaw of the target \gls{uav} is sampled uniformly at random:
\begin{align}
        \gamma^{(0)} &\sim U(0,2\pi) \\
    \mathcal{U}_d^{(t-1)} &\sim [U(-150,150),U(-150,150),U(200,300)]^T
\end{align}
where the initial pitch and roll is 0.

Subsequently, at every time step $t$, the radars' coordinate frames are transformed from the world coordinate frame to target coordinate frame: 
% via the homogeneous yaw, pitch, roll and translation projection inverse matrix multiplications:
\begin{align}
\begin{bmatrix}
    \Delta x ^{(t)}\\
    \Delta y ^{(t)}\\
    \Delta z ^{(t)}\\
\end{bmatrix}
= R_{\alpha^{(t)}}^{-1}  R_{\beta^{(t)}}^{-1} R_{\gamma^{(t)}}^{-1}(T^{(t)})^{-1} \mathcal{P} \label{eqn:w2t_coordinate} \\
\mathcal{P} = [p_1, p_2 \ldots p_J]
\end{align}
where $p_j$ is the world coordinates of  radar $j$. The azimuth and elevation corresponding to each \gls{rlos} are found by converting from Cartesian coordinates to Spherical coordinates:
\begin{align}
    \label{eqn:cart2spher}
    \rho^{(t)} = \sqrt{(\Delta x^{(t)})^2 + (\Delta y^{(t)})^2 + (\Delta z^{(t)})^2} \\
    \phi^{(t)} = \arctan \left(\frac{\Delta y^{(t)}}{\Delta x^{(t)}} \right) \\
        \theta^{(t)} = \delta(\Delta x^{(t)}<0)\arccos \left(\frac{-\Delta z^{(t)}}{\rho^{(t)}} \right) 
\end{align}
where $\rho$ is the length of the \gls{rlos} to the target \gls{uav}. \textcolor{black}{\cref{fig:yawpitchrolldiagram} visualizes the \gls{uav} coordinate frame highlighting the yaw, pitch, roll angles. The \cref{fig:yawpitchrolldiagram} also shows how the radar line of sight intersects with the target coordinate frame to determine the azimuth and elevation angle.}
A single time step of a simulated \gls{uav} being tracked by multiple individual radars is depicted in \cref{fig:simulated_rcs}, featuring the displayed azimuths and elevations. The simulated azimuth and elevation time series data, $\phi^{(1:t)} \in \mathcal{R}^{L \times J} \text{ and } \theta^{(1:t)} \in \mathcal{R}^{L \times J} $ respectively, is mapped to the measured \gls{rcs} time series data $z^{(1:t)} \in \mathcal{R}^{L \times (JF)}$. Due to the symmetry of the drones, the measured \gls{rcs} for any azimuth not between $[0,180]$ is approximated as the \gls{rcs} corresponding to $\phi=\phi+180$. We note a negative $\Delta z$ is used to align our target \gls{uav} coordinate frame with the experimental setup in \cite{rcs_data}, where the azimuth and elevation are calculated with respect to the bottom of the drone.

In summary, the test dataset $\mathcal{D}_{test}$ has the following structure:
\begin{align}
    \mathcal{D}_{test} = \{(z_i^{(1:L)},c_i)\}_{i=1}^{N_{test}}
\end{align}
where $z_i^{(t)}  = [x_{i1}^{(t)},x_{i2}^{(t)},\ldots,x_{iJ}^{(t)}]^T$ and $i$ denotes the sample index and $j$ the radar index.

We train the discriminative \gls{ml} model on the \gls{rcs}, azimuth, and elevation measurements described in \cref{sec:data_gen} \cref{sec:traindata}, and evaluate our \gls{ratr} framework on the trajectories described in \cref{sec:data_gen} \cref{sec:testdata}. We next discuss how these discriminative \gls{ml} models are integrated into the \gls{ratr} framework.

\vspace{-1em}
\section{METHODOLOGY}
\label{sec:methodology}
\begin{figure}[h!]
    \hspace{-1cm}
    \centering
    \includegraphics[width=9cm]{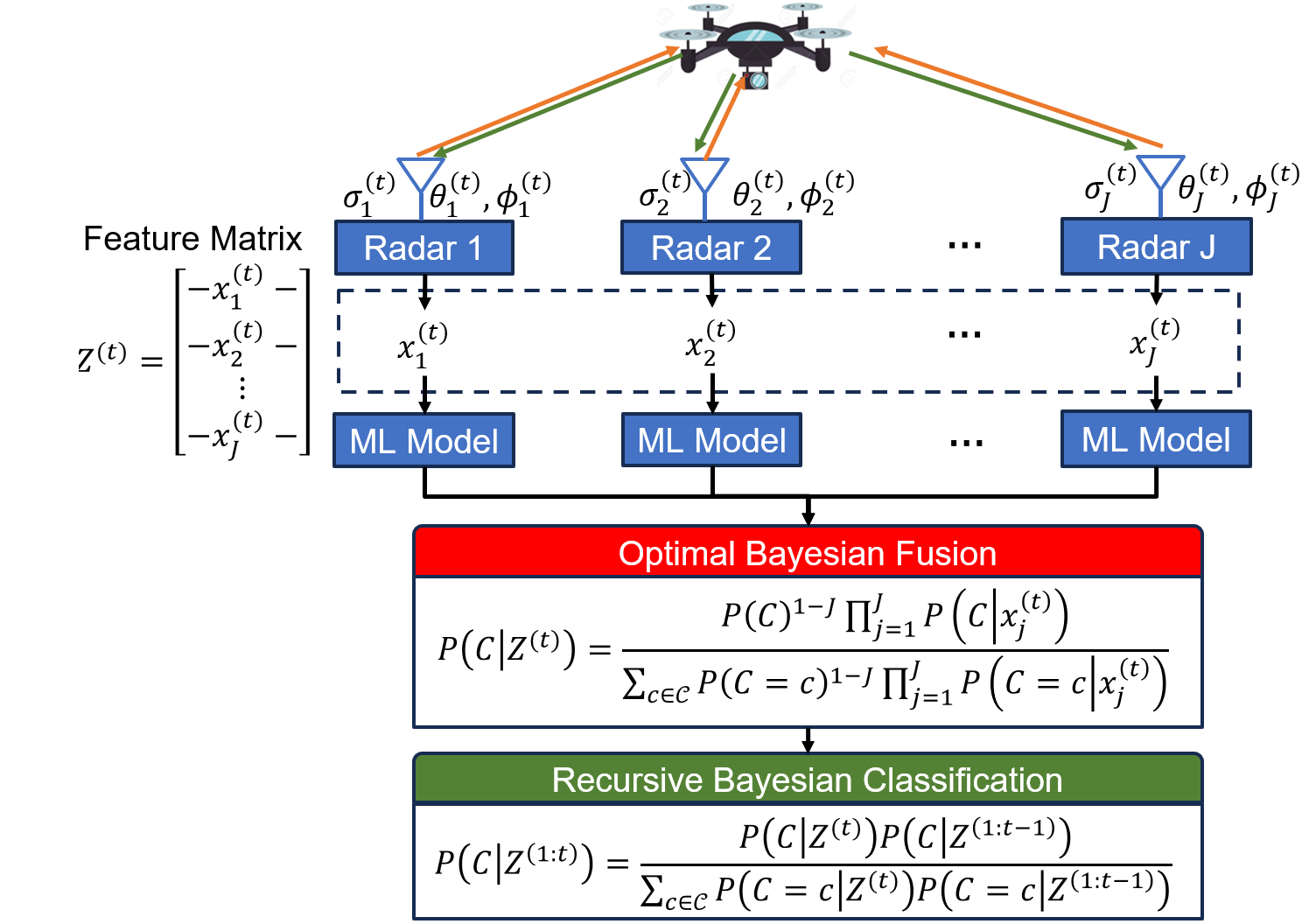}
    \caption{\small \gls{ratr} block diagram. Multiple radars continuously track and pulse an \gls{ew} at a moving \gls{uav} target. The each individual radar's belief about the \gls{uav} type is at time step $t$ is fused in an Bayesian way and used to update recursively a posterior belief about the \gls{uav} type. \textcolor{black}{In the experiments section, we replace the red block (the \gls{obf} fusion method) with non-Bayesian fusion methods for comparison, noting that any probabilistic fusion method which produces a probability distribution over drone type can replace the red block.}}
    \label{fig:atr_block_diagram}
    \vspace{-1em}
\end{figure}

At each time step $t$ multiple radars pulse \glspl{ew} at the target \gls{uav}, where each radar $j$ calculates a \gls{rcs} signature, azimuth, and elevation $x_j$. Each radar inputs its respective observation to a local discriminative \gls{ml} model to output a \gls{uav} type probability vector. Here, \textit{local} \textcolor{black}{denotes} that the \gls{ml} resides at the individual radar hardware. All the individual radar \gls{uav} type probability vectors are combined by a fusion method (\cref{sec:methodology} \cref{sec:fusion_method}). Subsequently, the posterior \gls{uav} type probability distribution, based on the past \gls{rcs} signatures $z^{(1:t-1)}$, is updated with the fused \gls{uav} type probability distribution from time step $t$. A \gls{rbc} framework is applied such that the posterior probability distribution recursively updates as the radars continuously track a moving \gls{uav}. Our framework is shown in \cref{fig:atr_block_diagram}.

We focus on three discriminative \gls{ml} models to generate individual radar \gls{uav} type probability vectors, which is described in the next subsection.

\begin{figure*}[ht!]
    \vspace{-1em}
    \centering
    \includegraphics[width=0.9\textwidth]{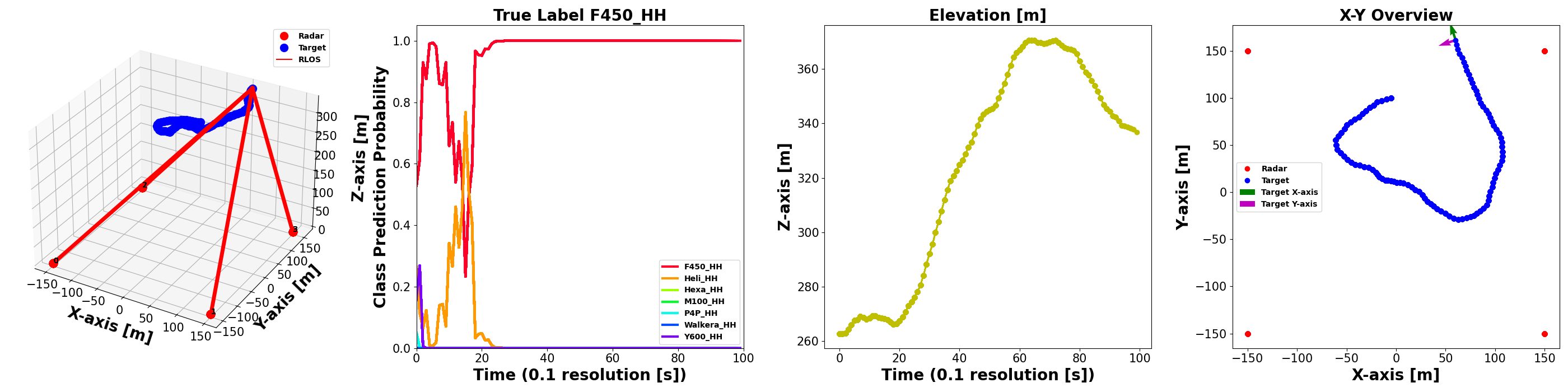}
    \caption{\small One \textcolor{black}{sample (a single trajectory for the F450 drone) from the test dataset} is shown for illustrative purposes. The first subfigure (from the left) depicts four uniformly spaced radars tracking and classifying a target \gls{uav} navigating with a random walk trajectory. The second subfigure illustrates the evolution of the \gls{rbc} posterior probability for the \gls{uav} type over time as more \gls{rcs}, azimuth, and elevation measurements are collected. The third and fourth figures display the change in elevation from ground level and the top view of the X-Y position of the \gls{uav} over time, respectively.}
    \label{fig:simulated_scenario}
    \vspace{-1em}
\end{figure*}

\vspace{-1em}
\subsection{ML Models}
In Logistic Regression, the model estimates the posterior probabilities of $|\mathcal{C}|$ classes parametrically through linear functions in  $x$. It adheres to the axioms of probability, providing a robust framework for probabilistic classification \cite{hastie2009elements}. The \gls{mlp}, on the other hand, is a parametric model which estimates a highly nonlinear function. We use a \gls{ffnn} architecture, where each neuron in the network involves an affine function of the preceding layer output, coupled with a nonlinear activation function (e.g., leaky-relu), with the exception of the last layer. For Logistic Regression and \gls{mlp}, we use the Python \textit{Scikit-learn} Python package \textit{LogisticRegression} and \textit{MLPClassifier} modules respectively; with the default hyperparameters (except the hidden layer size hyperparameter of [50,50,50]) \cite{sklearn}. \gls{xgboost} is a powerful non-parametric model that utilizes boosting with decision trees. In this ensemble method, each tree is fitted based on the residual errors of the previous trees, enhancing the model's overall predictive capability \cite{xgboost}. We use the Python \textit{XGBoost} package \textit{XGBClassifier} module  with the default hyperparameters \cite{xgboost}.

\vspace{-1em}
\subsection{Fusion Methods}
\label{sec:fusion_method}
The probability distribution $P(C|Z^{(t)})$ can leverage existing \gls{ml} discriminative classification algorithms, including but not limited to as \gls{xgboost}, \gls{mlp}, and Logistic Regression. For  each \gls{ml} model, we benchmark several radar \gls{rcs} fusion methodologies: \gls{obf}, average, and maximum.

\subsubsection{Bayesian Optimal Fusion}
The optimal fusion rule of the joint posterior distribution $p(C^{(t)}|Z^{(t)})$  is presented in \cite{optimalbayesianfusion,bishop2006pattern} as
\begin{align}
    P(C|Z^{(t)}) = \frac{P(C)^{(1-J)}\prod_{j=1}^J P(C|x_j^{(t)})}{\sum_{c=1}^\mathcal{C} P(C=c)^{(1-J)}\prod_{j=1}^J P(C=c|x_j^{(t)})}\label{eqn:bayesian_fuse}
\end{align}

The \gls{obf} method combines the individual radar probability vectors probabilistically, where the key assumption is conditional independence of the multiple individual radar observations at time step $t$ for a given \gls{uav} type: $P(x_1^{(t)},x_2^{(t)},\ldots, x_J^{(t)} | C=c) = \prod_{j=1}^J P(C|x_j^{(t)}) $ 

\subsubsection{Random}
We use random classification probability vectors for the fusion method as a baseline of comparison. We denote the fused classification probability of $J$ individual radars as a sample from the Dirichlet Distribution $P(C|Z^{(t)}) \sim Dir(\frac{1}{|\mathcal{C}|})$, which is equivalent to uniformly at random sampling from the simplex of discrete probability distributions.

\subsubsection{Hard-Voting}
Hard-voting involves selecting the \gls{uav} type that corresponds to the mode of decisions made by each individual radar discriminative \gls{ml} model regarding \gls{uav} types \cite{awe2024weighted}.
\begin{align}
    P(C=c|Z^{(t)}) =   \begin{cases}
   1-\frac{(|\mathcal{C}|-1)\epsilon}{|\mathcal{C}|}  & \text{if } c=Mo(\hat{c_1},\ldots,\hat{c_J})\\
   \frac{\epsilon}{|\mathcal{C}|}        & \text{otherwise}
\end{cases}
\end{align}
where $Mo$ denotes the mode function and $\hat{c_j}$ denotes the maximum probability \gls{uav} type for radar $j$.

\subsubsection{Soft-Voting}
Soft-voting is the average of multiple individual \gls{ml} model probability vectors. The \gls{uav} type with the highest average probability across all probability vectors is the final predicted \gls{uav} type. Averaging accounts for the confidence levels of each individual model, which provides a more nuanced method than hard-voting.
\begin{align}
    P(C|Z^{(t)})  =  \frac{1}{J} \sum_{j=1}^{J} P(C=c|x_{j}^{(t)}) \label{eqn:avg_fuse}
\end{align}

To benchmark our methodology, each individual radar operates within the same noisy channel (maintaining identical \gls{snr}) as they are closely geolocated. Consequently, the fusion method employed in this setup is analogous to the \gls{snr} fusion rule \cite{derham2010ambiguity}.

%the \gls{rbc} equation described next.
\subsubsection{Maximum}
We take the maximum ($\max$) class probability over multiple individual \gls{ml} models.
% Our intuition is that when the drone is directly above or near a radar, it should have higher classification accuracy because the bottom of the drone has the highest \gls{rcs} signature (due to the battery pack). However, due to the drone's ability to rapidly change its orientation over time, this previous assumption may not be correct. 
The max operator over random variables is known to be biased towards larger values, which will lead to worse classification performance \cite{van2016deep}.
\begin{align}
    P(C|Z^{(t)}) =  \max_{j} P(C=c|x_{j}^{(t)}) \label{eqn:max_fuse}
\end{align}

For each of the previously described fusion methods, when a new \gls{rcs} measurement at time $t$ is observed, the fused \gls{uav} type probability distribution updates the recursively estimated \gls{uav} type posterior distribution.

\begin{figure*}[ht!]
    \vspace{-1em}
    \centering
    \includegraphics[width=17cm]{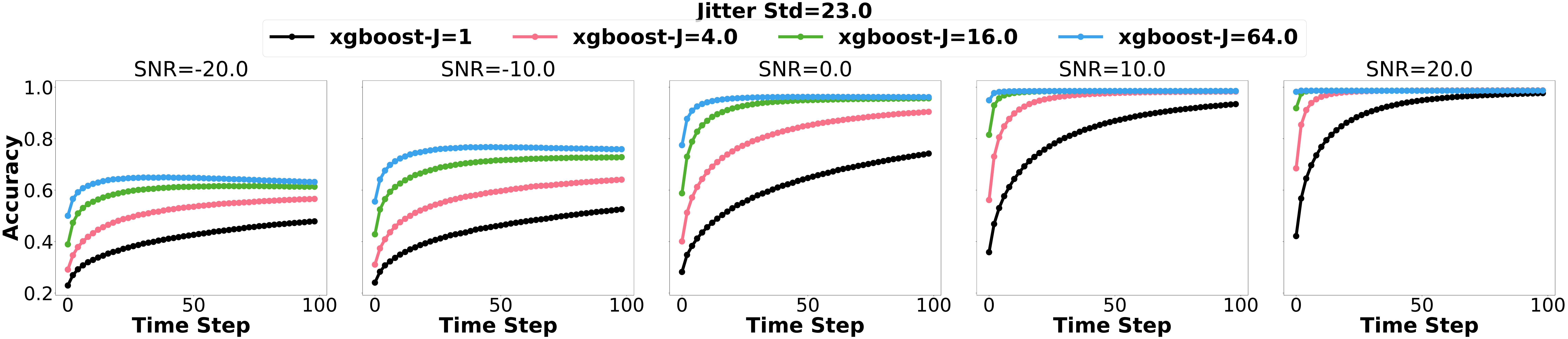}
    \caption{\small The \gls{rbc} accuracy performance \textcolor{black}{averaged across the test dataset} versus the number of radars increasing as a power of 2 (4, 16, 64 radars) for each discriminative \gls{ml} model. The subfigures from left to the right increase the \gls{snr} [dB] and the term $J$ in the legend denotes the number of radars in a surveillance area.}
    \label{fig:accuracy_vs_num_radar_snr}
\vspace{-1em}
\end{figure*}

\vspace{-1em}
\subsection{Recursive Bayesian Classification}
\gls{rbc} with the \gls{obf} provides a principled framework for Bayesian classification on time-series data. As  the radars' dwell time on an individual target increases, the uncertainty and accuracy in the \gls{uav} type will decrease and increase respectively due to increased number of data observations. Furthermore, Bayesian methods are robust to situations with noisy data, which is common in radar applications (high clutter environments, multi-scatter points, and multi-path propagation). The \gls{rbc} posterior probability for the \gls{uav} types is derived:

\begin{align}
    P(C|Z^{(1:t)}) \!\! &= \!\! \frac{P(Z^{(t)}|C)P(C|Z^{(1:t-1)})}{\!\sum_{c=1}^\mathcal{C}P(Z^{(t)}|C=c)P(C=c|Z^{(1:t-1)})} \nonumber\\
    \!\! &= \!\! \frac{\frac{P(C|Z^{(t)})\cancel{P(Z^{(t)})}}{\int P(C|Z^{(t)})P(Z^{(t)})}P(C|Z^{(1:t-1)})}{\!\sum_{c=1}^\mathcal{C}\frac{P(C=c|Z^{(t)})\cancel{P(Z^{(t)})}}{\int P(C=c|Z^{(t)})P(Z^{(t)})}P(C=c|Z^{(1:t-1)})} \nonumber\\
    \!\! &= \!\! \frac{P(C|Z^{(t)}) P(C|Z^{(1:t-1)})}{\!\sum_{c=1}^\mathcal{C} \! P(\!C=c|Z^{(t)}\!) \! P(\!C=c|Z^{(1:t-1)}\!)} \!
\end{align}

where we approximate the marginal distribution $P(C) = {\int P(C|Z^{(t)})P(Z^{(t)}) dZ^{(t)}}$ as a prior distribution, $\frac{1}{K}$, following previous papers \cite{calatrava2023recursive, Smedemark-Margulies_icassp_2023}.

Next, we outline our approach to benchmarking the integration of the \gls{obf} method with the \gls{rbc} within our \gls{ratr} framework, establishing a fully Bayesian target recognition approach.

\vspace{-1em}
\section{EXPERIMENTS}
\label{sec:experiment}
In the following subsections, we will describe the experimental setup and the assumptions made in our study to evaluate the performance of our proposed \gls{ratr} for \gls{uav} type classification. We perform 10 Monte Carlo trials for each experiment and report the respective average accuracy. Through comprehensive analysis, we elucidate the impact of varying \gls{snr} levels and fusion methodologies on \gls{ratr}.

\vspace{-1em}
\subsection{Configuration}
\label{sec:configuration}
All experiments were run on 
%Northeastern University's Discover Cluster short partition. The short partition workers have 
Dual Intel Xeon E5-2650@2GHz processors with 16 cores and 128 GB RAM. The training dataset and the testing dataset sizes are 10000 samples and 2000 samples respectively. We make the follow assumptions in our simulations of multistatic radar for single \gls{uav} type classification: \begin{enumerate}
    \item \textcolor{black}{for experimental purposes, we assume that the targets are always detected and the \gls{rcs} values are readily estimated (however, the methodology is independent of the detection procedure in radar pipelines)}.
    \item multiple radar systems transmit and receive independent signals (non-interfering constructively or destructively)  using time-division multiplexing,
    \item multiple radar systems continuously track a single \gls{uav} during a dwell time of 10 seconds (\cref{fig:simulated_scenario} leftmost figure),
    \item the azimuth and elevation of the incident \gls{ew} may be recovered with noise,
    \item use of \gls{acgn} for the \gls{rcs} signatures and uniform noise for the azimuth and elevations for noisy observations and
    \item the $J$ radars are uniformly spaced  in a 900m\textsuperscript{2} grid on the ground level (\cref{fig:simulated_scenario} leftmost figure).
\end{enumerate}

% we discussed this in previous meeting, does somehow have a citation that this is truly the optimal location of many radars when nothing is known about the target?

We analyze radar configurations under various scenarios, including uniform radar spacing and a single radar placed at the (-150 [m], -150 [m]) corner of the grid. Randomly positioning radars within an boxed-area is excluded to prevent redundant stochastic elements, given the inherent randomness in the drone trajectories. Our investigation focuses on radar fusion method performance across various \gls{snr} settings and the number of radars in the multistatic configuration.

\begin{figure*}[ht!]
    \vspace{-1em}
    \centering
    \includegraphics[width=0.85\linewidth]{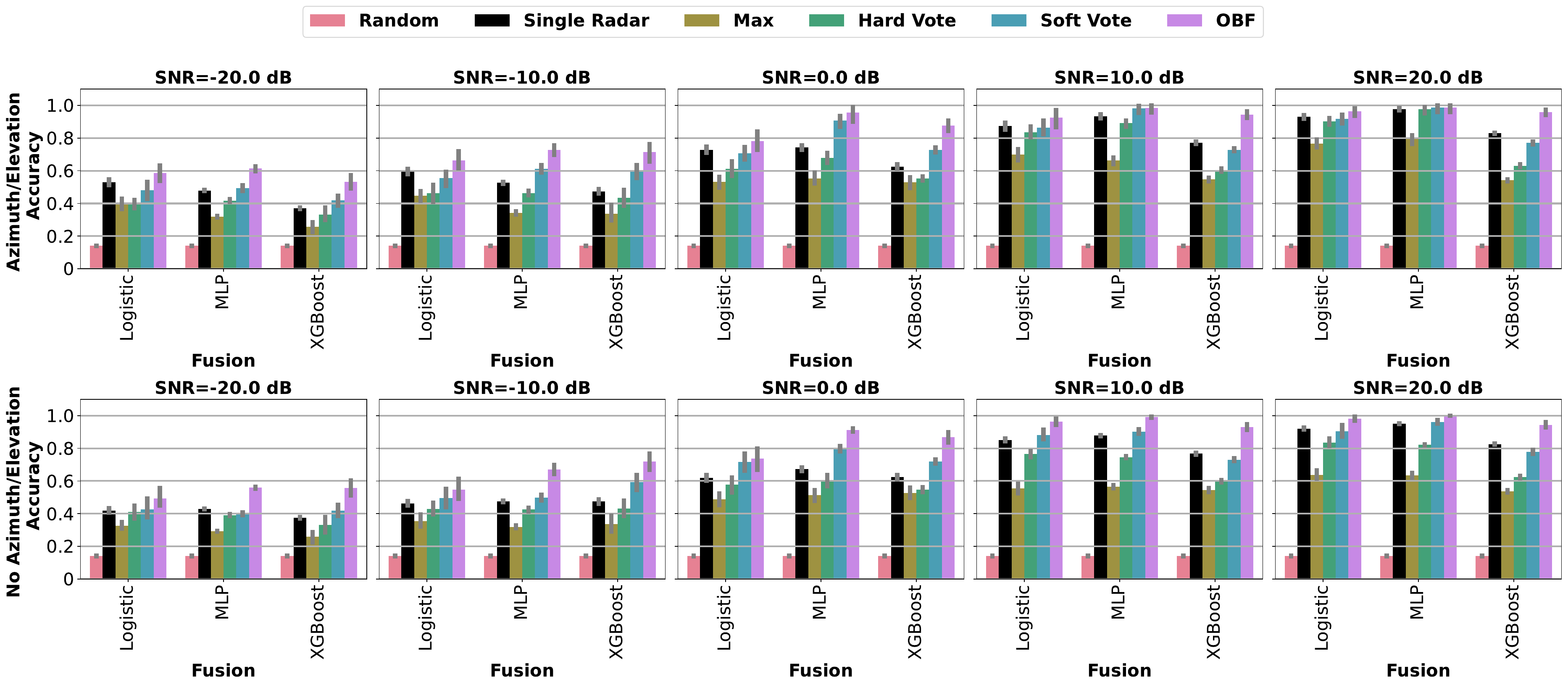}
    \caption{\small Each figure axis specifies the \gls{snr}, with the first row displaying results \textcolor{black}{averaged across the test dataset} with azimuth and elevation having a jitter standard deviation of 23 \textcolor{black}{degrees}, and the second row displaying results  \textcolor{black}{averaged across the test dataset} without azimuth and elevation \textcolor{black}{input features}. The bar color represents different fusion methods, and the line atop the bar indicates the 95\% confidence interval over 10 Monte Carlo trials. All results are for 16 radars in a multistatic configuration, except single radar.}
    \label{fig:accuracy_versus_snr}
    \vspace{-2em}
\end{figure*}

\vspace{-1em}
\subsection{Signal-to-Noise Ratio}
\label{sec:snr_detail}
We apply \gls{acgn} to the measured \gls{rcs} such as \cite{ezuma2021comparative}, but the units of the \gls{rcs} signatures are dBm\textsuperscript{2} units. The colored covariance matrix is generated as an outer product of a randomly sampled matrix of rank $F$:
\begin{align}
    M \sim \mathcal{N}(0,I_F) \\
    \Sigma = MM^T
\end{align}
For each observed \gls{rcs} signature $\sigma^*_i$ at time step $t$ and a specified \gls{snr}, we calculate the required noise power of the \gls{acgn}:
\begin{align}
    Tr(\Sigma_i) = \frac{\sum_f \sigma^*_i(f)^2}{10^{\frac{SNR}{10}}} \label{eqn:snr_vector}
\end{align}
where the trace of the covariance matrix is the total variance. We subsequently scale the covariance matrix to attain the desired \gls{snr} for sample $i$,
\begin{align}
\Sigma_i = \frac{\Sigma_i}{Tr(\Sigma_i)}\frac{\sum_f \sigma^*_i(f)^2}{10^{\frac{SNR}{10}}}
\end{align}
\begin{figure}[h!]
    % \vspace{-1em}
    \centering
    \includegraphics[width=7.7cm]{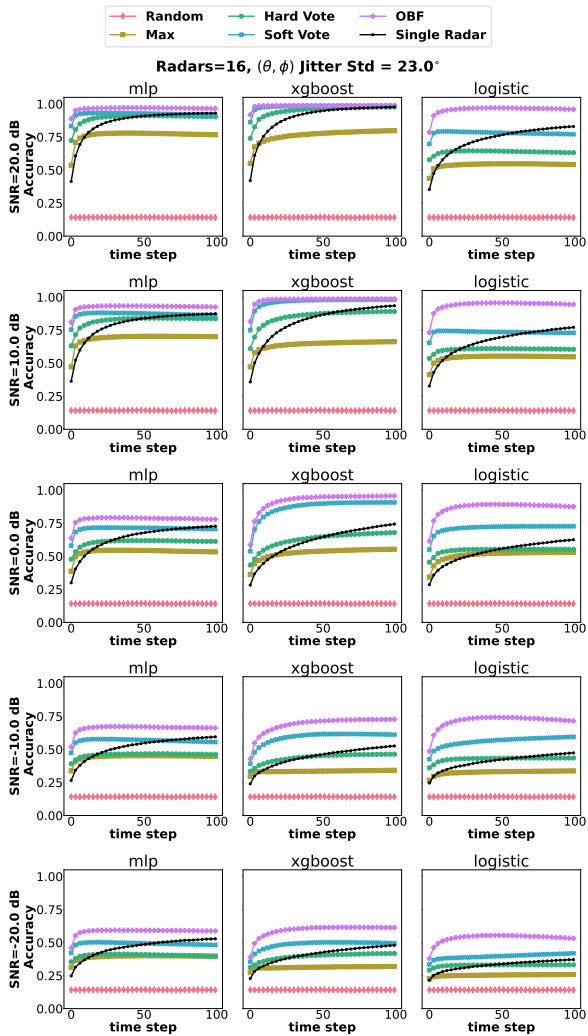}
    \caption{\small Each figure axis, where each row specifies the \gls{snr} and each column specifies the \gls{ml} model, plots the \gls{rbc} accuracy  \textcolor{black}{averaged across the test dataset} over time of 16 radars (and a single radar) for the \gls{obf}, soft-voting, hard-voting, single radar, and maximum fusion methods. The azimuth and elevation jitter standard deviations was $\approx$23 \textcolor{black}{degrees}. }
    \label{fig:accuracy_vs_snr_multi}
    \vspace{-2em}
\end{figure}
and then add a sample realization of the \gls{acgn} to the observed \gls{rcs} signature:
\begin{align}
    \sigma_i = \sigma_i^* + \epsilon && \epsilon \sim \mathcal{N}(0,\Sigma_i)
\end{align}
Additionally, to satisfy our third assumption of recovering the azimuth and elevation with noise, we add jitter uniformly at random to the ground truth azimuth and elevation with:
\begin{align}
    u_{\phi},u_{\theta} &\sim U_{\phi}(-a_{\phi},a_{\phi}),U_{\theta}(-a_{\theta},a_{\theta}) \\
    \begin{split}
    \phi= \phi^* + u_{\phi}
    \end{split}
    \begin{split}
    \theta = \theta^* + u_{\theta}
    \end{split}
\end{align}
where $a_\phi$ and $a_\theta$ determine the variance of uniform noise. We expect that our fully Bayesian \gls{ratr} for \gls{uav} type will be more robust to low \gls{snr} environments.

\vspace{-1em}
\subsection{Results}
Using the optimal fusion method, namely the \gls{obf} method, results in substantial classification performance improvement in multistatic radar configurations compared to a single monostatic radar configurations (\cref{fig:accuracy_vs_snr_multi},\cref{fig:accuracy_versus_snr}). 

However, when using a fusion method not rooted in Bayesian analysis, such as the common \gls{snr} fusion rule, a single radar may outperform or exhibit similar classification performance in multistatic radar configurations provided there is a sufficient dwell time (\cref{fig:accuracy_vs_snr_multi}, the column of \gls{mlp}  and the column of Logistic Regressionand \cref{fig:accuracy_versus_snr}). We also observed for the single radar and other fusion methods that various combinations of discriminative \gls{ml} models and \glspl{snr} plateau at substantially lower classification accuracy compared to the \gls{obf} (\cref{fig:accuracy_vs_snr_multi}). 

% \begin{table}[h!]
% \centering
% \scalebox{0.53}{
% \begin{tabular}{lrrrrrrrrrrrr}
% \toprule
% {} & \multicolumn{12}{c}{accuracy} \\
% ML Model & \multicolumn{4}{c}{logistic} & \multicolumn{4}{c}{mlp} & \multicolumn{4}{c}{xgboost} \\
% Fusion Method &   single &    max & average &    OBF & single &    max & average &    OBF &  single &    max & average &    OBF \\
% SNR   &          &        &         &        &        &        &         &        &         &        &         &        \\
% \midrule
%  20.0 &    0.825 &  0.537 &   0.779 &  0.944 &  0.930 &  0.735 &   0.958 &  0.987 &   0.951 &  0.634 &   0.962 &  0.999 \\
%  10.0 &    0.767 &  0.545 &   0.731 &  0.930 &  0.846 &  0.606 &   0.918 &  0.948 &   0.878 &  0.565 &   0.903 &  0.991 \\
%  0.0  &    0.623 &  0.528 &   0.720 &  0.868 &  0.622 &  0.521 &   0.707 &  0.698 &   0.672 &  0.513 &   0.799 &  0.912 \\
% -10.0 &    0.474 &  0.336 &   0.593 &  0.720 &  0.489 &  0.383 &   0.505 &  0.541 &   0.475 &  0.319 &   0.497 &  0.670 \\
% -20.0 &    0.374 &  0.260 &   0.419 &  0.557 &  0.406 &  0.326 &   0.485 &  0.438 &   0.428 &  0.292 &   0.400 &  0.559 \\
% \bottomrule
% \end{tabular}}
% \caption{\small The average accuracy for each fusion method and model combination is over 10 Monte Carlo trials, in a 16 multistatic radar configuration with azimuth and elevation jitter standard deviation of $\approx$23. The fusion method "single" refers to a monostatic radar configuration.}
% \label{tab:fusion_table}
% \vspace{-2em}
% \end{table}

\begin{figure*}[h!]
    \centering
    \includegraphics[width=12.5cm]{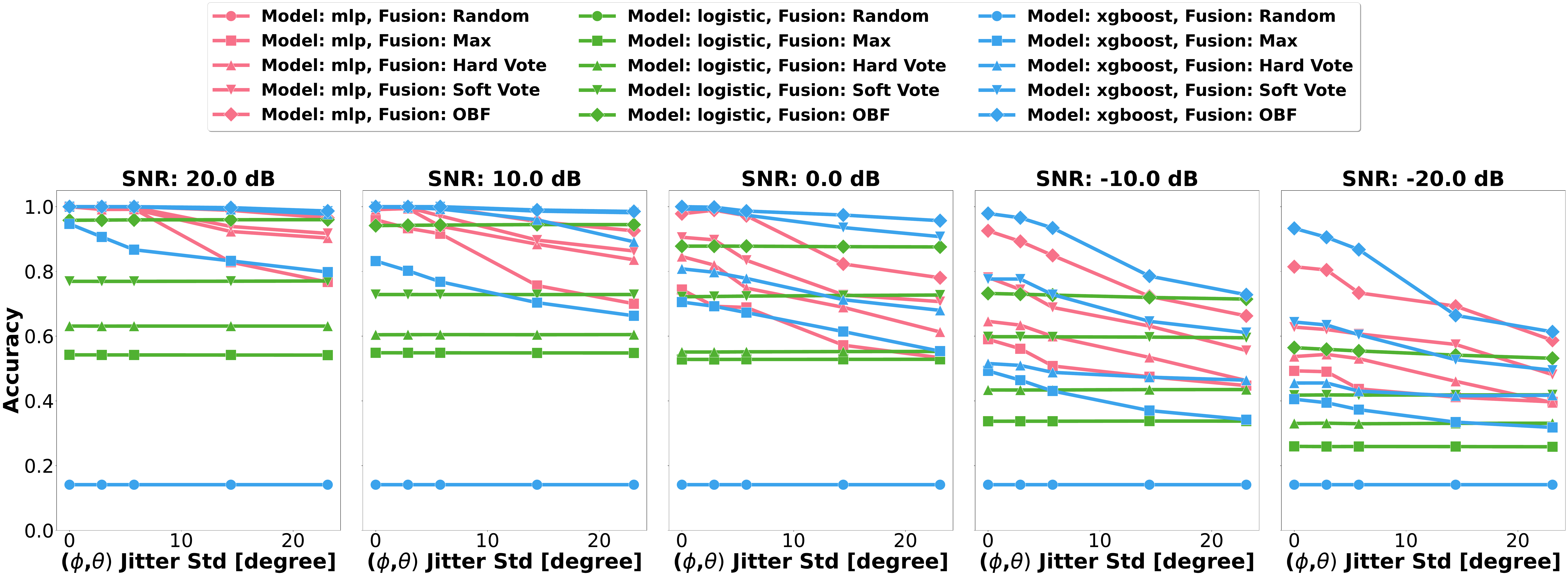}
    \caption{\small Each subfigure is the \gls{rbc} accuracy  averaged across the test dataset for 16 radars versus the azimuth and elevation standard deviation of the uniform distribution. The subfigures from left to the right decrease the \gls{snr} [dB]. Each line color denotes a different discriminative \gls{ml} model, and the line marker denotes the fusion method.}
    \label{fig:fusion_method}
    \vspace{-1em}
\end{figure*}

Increasing the number of radars within a surveillance area increases the different geometric views of the \gls{uav}, therefore providing a diverse \gls{uav} \gls{rcs} signature at a specific time step $t$. We see that as the number of radars increases, the classification performance also increases under various \gls{snr} environments (\cref{fig:accuracy_vs_num_radar_snr}). Furthermore, increasing the number of radars in a surveillance area allows correctly classifying the target \gls{uav} type in a shorter period of time (on average). 

We visualize the \gls{rbc} accuracy as the \gls{snr} changes (\cref{fig:accuracy_versus_snr}). We observe an ``S'' curve by following the top of the barplot across columnns of figure axes, where an increase in the signal power does not lead to an increase in classification performance at the upper extremes of \gls{snr}. However, radar channels typically operate in the 0 to -20 dB \gls{snr} range. When the \gls{ratr} is operating at lower \gls{snr} ranges, having more observations of the target \gls{uav} substantially improves the classification performance versus a single radar (\cref{fig:accuracy_vs_num_radar_snr}). 
If the \gls{snr} is high, the classification performance of a single radar should approach that of multi-radar classification, given a sufficiently large dwell time and suitable target geometries, especially when using a complex \gls{ml} model (\cref{fig:accuracy_vs_snr_multi}). 

% \begin{figure}[ht!]
%     \centering
%     \includegraphics[width=8cm]{images/accuracy_vs_snr.pdf}
%     \caption{\small The \gls{rbc} accuracy versus the \gls{snr} for each model and fusion method combination. The line color denotes a different discriminative \gls{ml} model, and the line marker denotes the fusion method.}
%     \label{fig:acc_vs_snr}
%     \vspace{-1em}
% \end{figure}

We assumed that azimuth and elevation of the incident \gls{ew} directed at the target \gls{uav} could be recovered for each radar, albeit with some noise. In the case of Logistic Regression, azimuth and elevation information does not seem to impact the classification decision. Increasing the standard deviation of the uniform noise jitter does not impact the classification accuracy. However, nonlinear discriminative \gls{ml} models, such as  \gls{mlp} and \gls{xgboost}, leverage the interactions between the \gls{rcs} signatures and the received azimuths and elevations, resulting in improved classification performance. Thus, for \gls{mlp} and \gls{xgboost}, we observe that an increase in the standard deviation of uniform noise for azimuth and elevation leads to a decrease on the classification performance for lower \gls{snr} environments (\cref{fig:fusion_method}).

As an ablation study, we completely remove the azimuth and elevation information from the discriminative \gls{ml} model training and inference. We observe that multistatic radar configurations still significantly improve \gls{rbc} classification performance, and the \gls{obf} method remains the most effective fusion method (\cref{fig:accuracy_versus_snr}).

% \begin{table}[h!]
% \centering
% \scalebox{0.53}{
% \begin{tabular}{l|rrrr|rrrr|rrrr}
% \toprule
%  & \multicolumn{12}{c}{Accuracy} \\
% % \midrule
% ML Model & \multicolumn{4}{c|}{logistic} & \multicolumn{4}{c|}{mlp} & \multicolumn{4}{c}{xgboost} \\
% Fusion Method &   single &    max & average &    OBF & single &    max & average &    OBF &  single &    max & average &    OBF \\
% SNR   &          &        &         &        &        &        &         &        &         &        &         &        \\
% \midrule
%  20.0 &    0.821 &  0.540 &   0.802 &  0.922 &  0.947 &  0.643 &   0.973 &  0.999 &   0.954 &  0.629 &   0.960 &  1.000 \\
%  10.0 &    0.757 &  0.552 &   0.740 &  0.888 &  0.875 &  0.560 &   0.930 &  0.925 &   0.884 &  0.580 &   0.921 &  0.996 \\
%  0.0  &    0.617 &  0.534 &   0.709 &  0.827 &  0.687 &  0.485 &   0.788 &  0.793 &   0.666 &  0.526 &   0.812 &  0.914 \\
% -10.0 &    0.468 &  0.343 &   0.575 &  0.701 &  0.504 &  0.381 &   0.535 &  0.621 &   0.455 &  0.312 &   0.487 &  0.656 \\
% -20.0 &    0.369 &  0.273 &   0.435 &  0.547 &  0.439 &  0.325 &   0.412 &  0.517 &   0.407 &  0.287 &   0.394 &  0.553 \\
% \bottomrule
% \end{tabular}}
% \caption{\small The average accuracy for each fusion method and model combination is computed over 10 Monte Carlo trials, considering a 16 multistatic radar configuration without azimuth and elevation information. The fusion method "single" refers to a monostatic radar configuration.}
% \label{tab:fusion_table_noazel}
% \vspace{-2em}
% \end{table}

\textcolor{black}{Although this paper focuses on the algorithmic perspective of \gls{uav} type classification, we note that without optimized code we achieve a 3.6 milliseconds (ms) inference time at a given time step on the CPU of a Zephyrus M16 laptop with a 12th Gen Intel(R) Core(TM) i9-12900H 2.5 GHz processor. That is, in 3.6 ms each local model outputs a probability distribution, the local model probability distributions are fused with the \gls{obf} method, and the running global posterior prediction is updated with \gls{rbc}.}

\vspace{-1em}
\section{CONCLUSION}
This paper is the first to introduce a fully Bayesian \gls{ratr} for \gls{uav} type classification in multistatic radar configurations using \gls{rcs} signatures. We evaluated the classification accuracy and robustness of our method across diverse \gls{snr} settings using \gls{rcs}, azimuth, and elevation time series data generated by random walk drone trajectories. Our empirical results demonstrate that integrating the \gls{obf} method  with \gls{rbc} in multistatic radar \textcolor{black}{significantly} enhances \gls{atr}. Additionally, our method exhibits greater robustness in lower \gls{snr}, with larger relative improvements observed compared to single monostatic radar and other non-trivial fusion methods. Thus, fully Bayesian \gls{ratr} in multistatic radar configurations using \gls{rcs} signatures improves both classification accuracy and robustness.

Future work will integrate radar-based domain-expert knowledge with Bayesian analysis, such that our method may incorporate radar parameters such as the range resolution or estimated \gls{snr}.

% GS: Acknowledgement goes up front for this journal, 
% I just noticed when disabling other content
%\section*{ACKNOWLEDGMENT}
\vspace{-1em}
\bibliography{bibliography}

\begin{IEEEbiography}[{\includegraphics[width=0.6in,height=1.25in,clip,keepaspectratio]{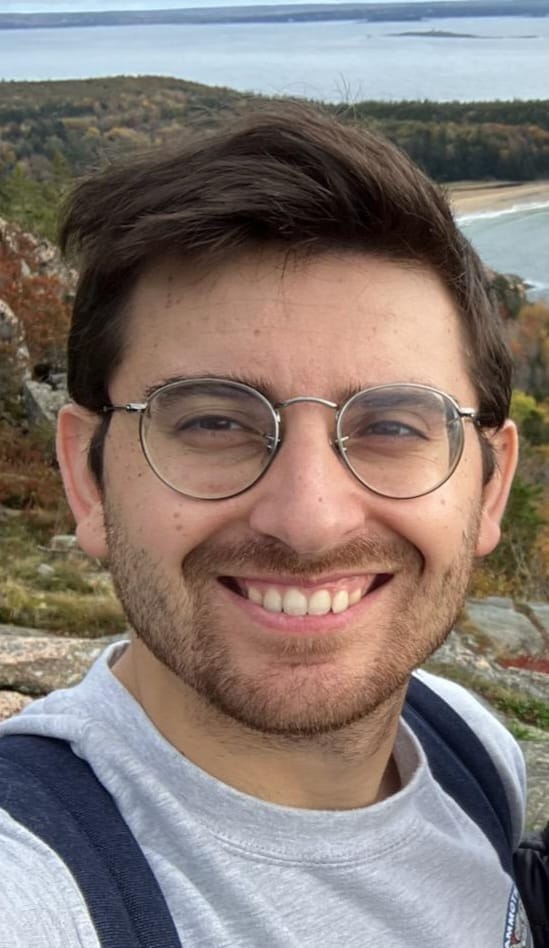}}]{Michael Potter} is a Ph.D. student at Northeastern University under the advisement of Deniz Erdo\u{g}mu\c{s} of the Cognitive Systems Laboratory (CSL). He received his B.S, M.S., and M.S.  degrees in Electrical and Computer Engineering from Northeastern University and University of California Los Angeles (UCLA) in 2020, 2020, and 2022 respectively. His research interests are in recommendation systems, Bayesian Neural Networks, uncertainty quantification, and dynamics based manifold learning. 
\end{IEEEbiography}%
\vspace{-1em}

\begin{IEEEbiography}[{\includegraphics[width=1in,height=1.25in,clip,keepaspectratio]{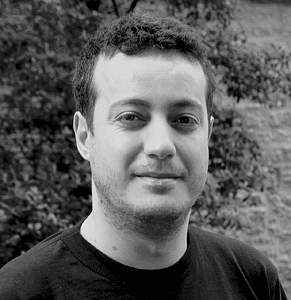}}]%
{Murat Akcakaya} (Senior Member, IEEE)
received his Ph.D. degree in Electrical Engineering from the Washington University in St. Louis, MO, USA, in December 2010. He is currently an Associate Professor in the Electrical and Computer Engineering Department of the University of Pittsburgh. His research interests are in the areas of statistical signal processing and machine learning.
\end{IEEEbiography}
\vspace{-1em}

\begin{IEEEbiography}[{\includegraphics[width=1in,height=1.25in,clip,keepaspectratio]{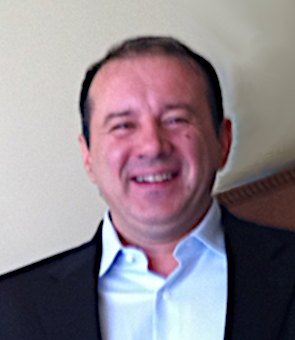}}]
{Marius Necsoiu} (Member, IEEE) {\space}received the MS in EE (Applied Electronics) in 1992 from the Politehnica University of Bucharest, Romania and PhD in Environmental Science (Remote Sensing) in 2000 from the University of North Texas. He has broad experience and expertise in remote sensing systems, radar data modeling, and analysis to characterize electromagnetic environment behavior and geophysical deformation. As part of the DEVCOM ARL he leads research in cognitive radars and explores new paradigms in AI/ML science that are applicable in radar/EW research.
\end{IEEEbiography}
\vspace{-1em}

\begin{IEEEbiography}
[{\includegraphics[width=1in,height=1.25in,clip,keepaspectratio]{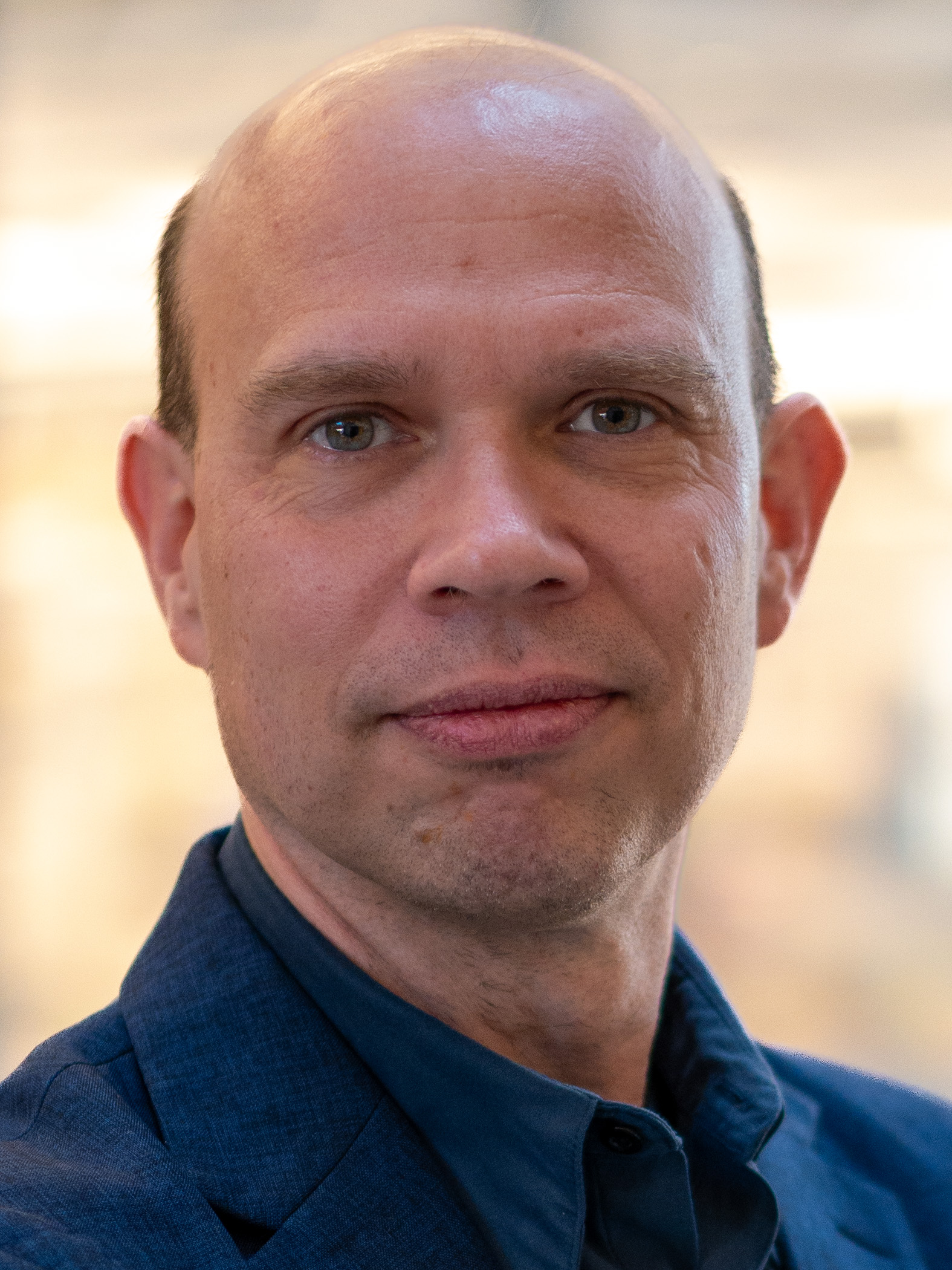}}]
{Gunar Schirner} (S'04--M'08) holds PhD (2008) and MS (2005) degrees in
  electrical and computer engineering from the University of California,
  Irvine. He is currently an Associate Professor in Electrical and
  Computer Engineering at Northeastern University. His research interests
  include the modelling and design automation principles for domain
  platforms, real-time cyber-physical systems and the
  algorithm/architecture co-design of high-performance efficient edge
  compute systems.
\end{IEEEbiography}
\vspace{-1em}

\begin{IEEEbiography}
[{\includegraphics[width=1in,height=1.25in,clip,keepaspectratio]{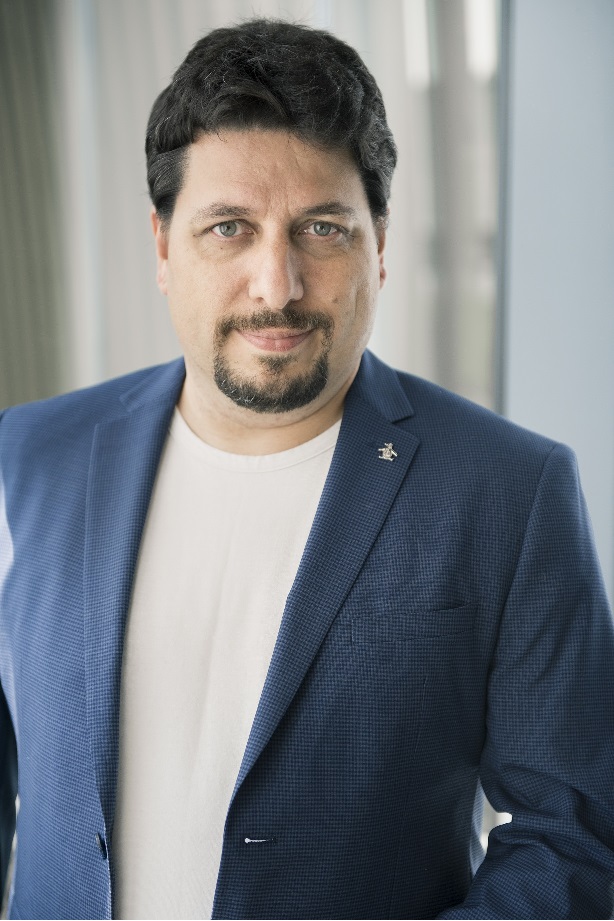}}]
{Deniz Erdo\u{g}mu\c{s}} (Sr Member, IEEE), received BS in EE and Mathematics (1997), and MS in EE (1999) from the Middle East Technical University, PhD in ECE (2002) from the University of Florida, where he was a postdoc until 2004. He was with CSEE and BME Departments at OHSU (2004-2008). Since 2008, he has been with the ECE Department at Northeastern University. His research focuses on statistical signal processing and machine learning with applications data analysis, human-cyber-physical systems, sensor fusion and intent inference for autonomy. He has served as associate editor and technical committee member for multiple IEEE societies.
\end{IEEEbiography}
\vspace{-1em}

\begin{IEEEbiography}
[{\includegraphics[width=1in,height=1.25in,clip,keepaspectratio]{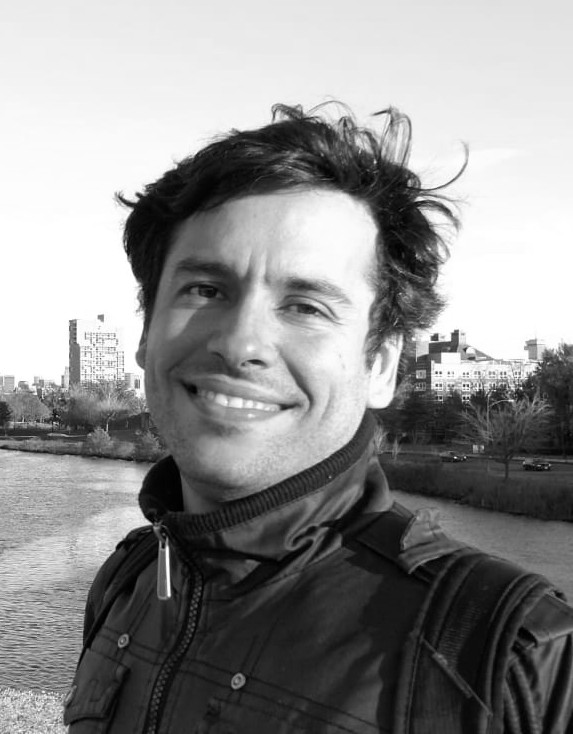}}]%{figures/carnival_floripa_tales}}]
{Tales Imbiriba} (Member, IEEE)   
is an Assistant Research Professor at the ECE dept., and Senior Research Scientist at the Institute for Experiential AI, both at Northeastern University (NU), Boston, MA, USA. He
received his Doctorate degree from the Department of Electrical Engineering (DEE) of the Federal University of Santa Catarina (UFSC), Florian\'opolis, Brazil, in 2016. He served as a Postdoctoral Researcher at the DEE--UFSC (2017--2019) and at the ECE dept. of the NU (2019--2021). 
His research interests include audio and image processing, pattern recognition, Bayesian inference, online learning, and physics-guided machine learning.
% \end{IEEEbiographynophoto}
\end{IEEEbiography}
\vspace{-1em}

\end{document}